\documentclass[runningheads]{llncs}
\usepackage[T1]{fontenc}
\usepackage{graphicx}
\usepackage{booktabs}
\usepackage[colorlinks,
            linkcolor=blue,
            anchorcolor=blue,
            citecolor=blue]{hyperref} % 颜色可调
\usepackage{multirow}
\usepackage{amssymb}
\usepackage{marvosym}

\begin{document}
\title{Shifting More Attention to Breast Lesion Segmentation in Ultrasound Videos}
\author{
Junhao Lin      \inst{1} \and
Qian Dai        \inst{1} \and
Lei Zhu         \inst{2,3,} $^\textrm{\Letter}$ \and
Huazhu Fu       \inst{4} \and
Qiong Wang      \inst{5} \and
Weibin Li       \inst{6} \and
Wenhao Rao      \inst{1} \and
Xiaoyang Huang  \inst{1,}$^\textrm{\Letter}$ \and
Liansheng Wang  \inst{1} 
}
%index{Junhao, Lin}
%index{Qian, Dai}
%index{Lei, Zhu}
%index{Huazhu, Fu}
%index{Qiong, Wang}
%index{Weibin, Li}
%index{Xiaoyang, Huang}
%index{Liansheng, Wang}

\authorrunning{Junhao Lin et al.}

\institute{
School of Informatics, Xiamen University, Xiamen, China \\
\email{xyhuang@xmu.edu.cn}
\and
ROAS Thrust, System Hub, The Hong Kong University of Science and Technology (Guangzhou), Guangzhou, China
\\ \email{leizhu@ust.hk}
\and
Department of Electronic and Computer Engineering, The Hong Kong University of Science and Technology, Hong Kong SAR, China
\and
Institute of High Performance Computing, Agency for Science, Technology and Research, Singapore
\and
Guangdong Provincial Key Laboratory of Computer Vision and Virtual Reality, Shenzhen Institute of Advanced Technology, Chinese Academy of Sciences, Shenzhen, China
\and
School of Medicine, Xiamen University, Xiamen, China
}
\maketitle              % typeset the header of the contribution
\begin{abstract}
Breast lesion segmentation in ultrasound (US) videos is essential for diagnosing and treating axillary lymph node metastasis.
However, the lack of a well-established and large-scale ultrasound video dataset with high-quality annotations has posed a persistent challenge for the research community. To overcome this issue, we meticulously curated a US video breast lesion segmentation dataset comprising 572 videos and 34,300 annotated frames, covering a wide range of realistic clinical scenarios.
Furthermore, we propose a novel frequency and localization feature aggregation network (FLA-Net) that learns temporal features from the frequency domain and predicts additional lesion location positions to assist with breast lesion segmentation. We also devise a localization-based contrastive loss to reduce the lesion location distance between neighboring video frames within the same video and enlarge the location distances between frames from different ultrasound videos.
Our experiments on our annotated dataset and two public video polyp segmentation datasets demonstrate that our proposed FLA-Net achieves state-of-the-art performance in breast lesion segmentation in US videos and video polyp segmentation while significantly reducing time and space complexity.
Our model and dataset are available at \href{https://github.com/jhl-Det/FLA-Net}{https://github.com/jhl-Det/FLA-Net}.

\end{abstract}

% #############################
\section{Introduction}
Axillary lymph node (ALN) metastasis is a severe complication of cancer that can have devastating consequences, including significant morbidity and mortality. Early detection and timely treatment are crucial for improving outcomes and reducing the risk of recurrence.
In breast cancer diagnosis, accurately segmenting breast lesions in ultrasound (US) videos is an essential step for computer-aided diagnosis systems, as well as breast cancer diagnosis and treatment. However, this task is challenging due to several factors, including blurry lesion boundaries, inhomogeneous distributions, diverse motion patterns, and dynamic changes in lesion sizes over time~\cite{cvanet}.

\begin{table}[!t]
\caption{
 Statistics of existing breast lesion US videos datasets and the proposed dataset.
 \textbf{\#videos}: numbers of videos.
 \textbf{\#AD}: number of annotated frames.
 \textbf{BBox}: whether provide bounding box annotation.
 \textbf{BBox}: whether provide segmentation mask annotation.
 \textbf{BM}: whether provide lesion classification label (Benign or Malignant).
 \textbf{PA}: whether provide axillary lymph node (ALN) metastasis label (Presence or Absence).
}
\label{tab:datasets-cmp}
% \vspace{-3mm}
\centering
\begin{tabular}{l|ccc|cccc}
\hline
  Dataset & Year & \#videos & \#AF & BBox & Mask &BM & PA\\
\hline
 Li et al.~\cite{RethinkingLiJialu} & 2022 & 63 & 4,619 & $\times$    & $\checkmark$ & $\checkmark$  & $\times$  \\
 Lin et al.~\cite{cvanet} & 2022 & 188 & 25,272 & $\checkmark$ & $\times$ & $\checkmark$ & $\times$ \\

\hline
 Ours & 2023 & \textbf{572} & \textbf{34,300} & $\checkmark$    & $\checkmark$ & $\checkmark$ & $\checkmark$ \\
  
\hline
\end{tabular}
\vskip -5pt
\end{table}

The work presented in \cite{RethinkingLiJialu} proposed the first pixel-wise annotated benchmark dataset for breast lesion segmentation in US videos, but it has some limitations. Although their efforts were commendable, this dataset is private and contains only 63 videos with 4,619 annotated frames. The small dataset size increases the risk of overfitting and limits the generalizability capability. 
In this work, \textbf{we collected a larger-scale  US video breast lesion segmentation  dataset} with 572 videos and 34,300 annotated frames, of which 222 videos contain ALN metastasis, covering a wide range of realistic clinical scenarios.
Please refer to Table~\ref{tab:datasets-cmp} for a detailed comparison between our dataset and existing datasets.

Although the existing benchmark method DPSTT~\cite{RethinkingLiJialu} has shown promising results for breast lesion segmentation in US videos, it only uses the ultrasound image to read memory for learning temporal features. However, ultrasound images suffer from speckle noise, weak boundaries, and low image quality. Thus, there is still considerable room for improvement in ultrasound video breast lesion segmentation. 
To address this, \textbf{we propose a novel network called Frequency and Localization Feature Aggregation Network (FLA-Net)} to improve breast lesion segmentation in ultrasound videos. Our FLA-Net learns frequency-based temporal features and then uses them to predict auxiliary breast lesion location maps to assist the segmentation of breast lesions in video frames.
Additionally, we devise a contrastive loss to enhance the breast lesion location similarity of video frames within the same ultrasound video and to prohibit location similarity of different ultrasound videos.
The experimental results unequivocally showcase that our network surpasses state-of-the-art techniques in the realm of both breast lesion segmentation in US videos and two video polyp segmentation benchmark datasets.

\section{Ultrasound Video  Breast Lesion Segmentation Dataset}

\begin{figure*}[!t]
\centering
\includegraphics[width=\textwidth]{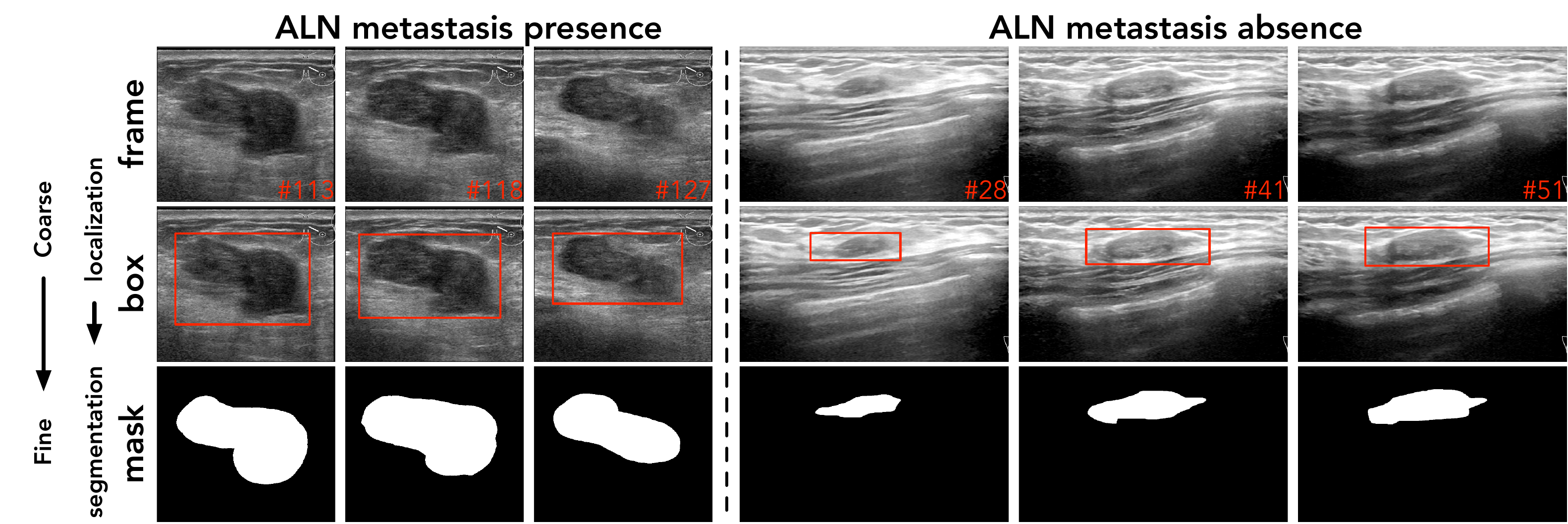}
\vspace{-5mm}
\caption{Examples of our ultrasound video dataset for breast lesion segmentation.} 

\label{fig:dataset}
\end{figure*}

To support advancements in breast lesion segmentation and ALN metastasis prediction, we collected a dataset containing 572 breast lesion ultrasound videos with 34,300 annotated frames. Table~\ref{tab:datasets-cmp} summarizes the statistics of existing breast lesion US video datasets.
Among 572 videos, 222 videos with ALN metastasis.
Nine experienced pathologists were invited to manually annotate breast lesions at each video frame. 
Unlike previous datasets~\cite{RethinkingLiJialu,cvanet}, our dataset has a reserved validation set to avoid model overfitting.
The entire dataset is partitioned into training, validation, and test sets in a proportion of 4:2:4, yielding a total of 230 training videos, 112 validation videos, and 230 test videos for comprehensive benchmarking purposes.
Moreover, apart from the segmentation annotation, our dataset also includes lesion bounding box labels, which enables benchmarking breast lesion detection in ultrasound videos.
More dataset statistics are available in the \textit{Supplementary}.

% #############################
\section{Proposed Method}
\begin{figure*}[!t]
\centering
\includegraphics[width=0.90\textwidth]{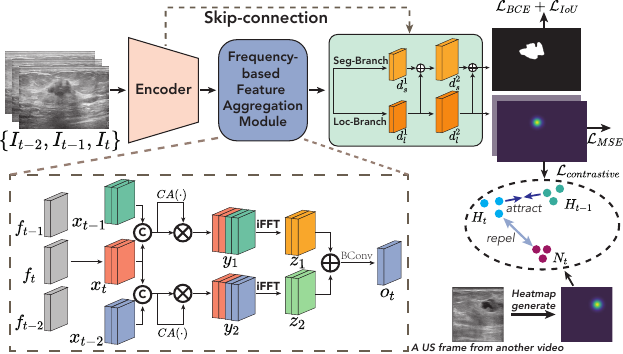}
\caption{Overview of our FLA-Net.
Our network takes an ultrasound frame $I_t$ and its adjacent two frames ($I_{t-1}$ and $I_{t-2}$) as input. 
Three frames are first passed through an encoder to learn three CNN features ($f_{t}$, $f_{t-1}$, and $f_{t-2}$). 
Then Frequency-based Feature Aggregation Module is then used to aggregate these features and the aggregated feature map is then passed into our two-branch decoder to predict the breast lesion segmentation mask of $I_t$, and a lesion localization heatmap. 
Moreover, we devise a location-aware contrastive loss (see $\mathcal{L}_{contrastive}$) to reduce location distance of frames from the same video and enlarge the location distance of different video frames.
}
\label{fig:framework}
\vspace{-5pt}
\end{figure*}

Figure~\ref{fig:framework} provides a detailed illustration of the proposed frequency and localization feature aggregation network (FLA-Net).
When presented with an ultrasound frame denoted as $I_t$ along with its two adjacent video frames ($I_{t-1}$ and $I_{t-2}$), our initial step involves feeding them through an Encoder, specifically the Res2Net50 architecture~\cite{res2net}, to acquire three distinct features labeled as $f_t$, $f_{t-1}$, and $f_{t-2}$.
Then, we devise a frequency-based feature aggregation (FFA) module to integrate freqency features of each video frame.
After that, we pass the output features $o_{t}$ of the FFA module into two decoder branches (similar to the UNet decoder~\cite{unet}): one is the localization branch to predict the localization map of the breast lesions, while another segmentation branch integrates the features of the localization branch to fuse localization feature for segmenting breast lesions.
Moreover, we devise a location-based contrastive loss to regularize the breast lesion locations of inter-video frames and intra-video frames.

\subsection{Frequency-based Feature Aggregation (FFA) Module}
According to the spectral convolution theorem in Fourier theory, any modification made to a single value in the spectral domain has a global impact on all the original input features~\cite{bergland1969guided}. 
This theorem guides the design of FFA module, which has a global receptive field to refine features in the spectral domain.
As shown in Fig.~\ref{fig:framework}, our FFA block takes three features ($f_{t} \in \mathbb{R}^{c\times h \times w} $, $f_{t-1} \in \mathbb{R}^{c\times h \times w} $, and $f_{t-2} \in \mathbb{R}^{c\times h \times w}$) as input.
To integrate the three input features and extract relevant information while suppressing irrelevant information, our FFA block first employs a Fast Fourier Transform (FFT) to transform the three input features into the spectral domain, resulting in three corresponding spectral domain features ($\hat{f}_{t} \in \mathbb{C}^{c\times h \times w} $, $\hat{f}_{t-1} \in \mathbb{C}^{c\times h \times w} $, and $\hat{f}_{t-2} \in \mathbb{C}^{c\times h \times w}$), which capture the frequency information of the input features.
%%%%%%%%%%%%%%%%%%%
Note that the current spectral features ($\hat{f}_{t}$,$\hat{f}_{t-1}$, and $\hat{f}_{t-2}$) are complex numbers and incompatible with the neural layers.
Therefore we concatenate the real and imaginary parts of these complex numbers along the channel dimension respectively and thus obtain three new tensors ($x_{t} \in \mathbb{R}^{2c\times h \times w}$, $x_{t-1} \in \mathbb{R}^{2c\times h \times w}$, and $x_{t-2} \in \mathbb{R}^{2c\times h \times w}$) with double channels.
%%%%%%%%%%%%%%%%%%%
Afterward, we take the current frame spectral-domain features $x_{t}$ as the core and fuse the spatial-temporal information from the two auxiliary spectral-domain features ($x_{t-1}$ and $x_{t-2}$), respectively.
Specifically, we first group three features into two groups ($\{x_{t}, x_{t-1}\}$ and $\{x_{t}, x_{t-2}\}$) and develop a channel attention function $CA(\cdot)$ to obtain two attention maps.
The $CA(\cdot)$ passes an input feature map to a feature normalization, two 1$\times$1 convolution layers $Conv(\cdot)$, a ReLU activation function $\delta(\cdot)$, and a sigmoid function $\sigma(\cdot)$ to compute an attention map.
Then, we element-wise multiply the obtained attention map from each group with the input features, and the multiplication results (see $y_1$ and $y_2$) are then transformed into complex numbers by splitting them into real and imaginary parts along the channel dimension.
After that, inverse FFT (iFFT) operation is employed to transfer the spectral features back to the spatial domain, and then two obtained features at the spatial domain are denoted as $z_1$ and $z_2$.
Finally, we further element-wisely add $z_1$ and $z_2$ and then pass it into a \textit{``$BConv$''} layer to obtain the output feature $o_t$ of our FFA module.
Mathematically, $o_t$ is computed by ${o}_t = BConv(z_1 + z_2)$, where \textit{``$BConv$''} contains a $3\times3$ convolution layer, a group normalization, and a $ReLU$ activation function.

\subsection{Two-branch Decoder}
After obtaining the frequency features, we introduce a two-branch decoder consisting of a segmentation branch and a localization branch to incorporate temporal features from nearby frames into the current frame. 
Each branch is built based on the UNet decoder~\cite{unet} with four convolutional layers.
Let $d_s^1$ and $d_s^2$ denote the features at the last two layers of the segmentation decoder branch, and $d_l^1$ and $d_l^2$ denote the features at the last two layers of the localization decoder branch.
Then, we pass $d_l^1$ at the localization decoder branch to predict a breast lesion localization map.
Then, we element-wisely add $d_l^1$ and $d_s^1$, and element-wisely add $d_l^2$ and $d_s^2$, and pass the addition result into a ``$BConv$'' convolution layer to predict the segmentation map $S_t$ of the input video frame $I_t$.

\vspace{.5mm}
\noindent
\textbf{Location Ground Truth}.
Instead of formulating it as a regression problem, we adopt a likelihood heatmap-based approach to encode the location of breast lesions, since it is more robust to occlusion and motion blur. 
To do so, we compute a bounding box of the annotated breast lesion segmentation result, and then take the center coordinates of the bounding box.
After that, we apply a Gaussian kernel with a standard deviation of 5 on the center coordinates to generate a heatmap, which is taken as the ground truth of the breast lesion localization.

\subsection{Location-based Contrastive Loss}
Note that the breast lesion locations of neighboring ultrasound video frames are close, while the breast lesion location distance is large for different ultrasound videos, which are often obtained from different patients. 
Motivated by this, we further devise a location-based contrastive loss to make the breast lesion locations at the same video to be close, while pushing the lesion locations of frames from different videos away. By doing so, we can enhance the breast lesion location prediction in the localization branch.
Hence, we devise a \textbf{location-based contrastive loss} based on a triplet loss~\cite{triplet_loss}, and the definition is given by: 
\begin{equation}
\mathcal{L}_{contrastive} = max(MSE(H_{t}, H_{t-1}) - MSE(H_{t}, N_{t}) + \alpha, 0),
\end{equation}
where $\alpha$ is a margin that is enforced between positive and negative pairs. $H_t$ and $H_{t-1}$ are predicted heatmaps of neighboring frames from the same video. $N_t$ denotes the heatmap of the breast lesion from a frame from another ultrasound video.
Hence, the total loss $\mathcal{L}_{total}$ of our network is computed by:
\begin{equation} \small
\mathcal{L}_{total} = \mathcal{L}_{contrastive} + \lambda_1\mathcal{L}_{MSE} (H_{t}, G^H_{t}) + \lambda_2 \mathcal{L}_{BCE} (S_{t}, G^S_{t}) + \lambda_3 \mathcal{L}_{IoU} (S_{t}, G^S_{t}) ,
\end{equation}
where $G^H_{t}$ and $G^S_{t}$ denote the ground truth of the breast lesion segmentation and the breast lesion localization. We empirically set weights $\lambda_1$=$\lambda_2$=$\lambda_3$=1.

% #############################
\section{Experiments and Results}
\noindent
\textbf{Implementation Details.}
To initialize the backbone of our network, we pretrained Res2Net-50~\cite{res2net} on the ImageNet dataset, while the remaining components of our network were trained from scratch.
Prior to inputting the training video frames into the network, we resize them to $352 \times 352$ dimensions.
Our network is implemented in PyTorch and employs the Adam optimizer with a learning rate of $5 \times 10^{-5}$, trained over 100 epochs, and a batch size of 24.
Training is conducted on four GeForce RTX 2080 Ti GPUs.
For quantitative comparisons, we utilize various metrics, including the Dice similarity coefficient (Dice), Jaccard similarity coefficient (Jaccard), F1-score, and mean absolute error (MAE).

%%%%%%%%%%%%%%%%%%%%%%%%%%%%
\begin{table}[!t]
\caption{
 Quantitative comparisons between our FLA-Net and the state-of-the-art methods on our test set in terms of breast lesion segmentation in ultrasound videos.
}
\vspace{-3mm}
\label{tab:sota}
\centering
\begin{tabular}{l|c|c|c|c|c} 
  \hline
  Method &image/video & Dice $\uparrow$ & Jaccard $\uparrow$ & F1-score $\uparrow$& MAE $\downarrow$ \\
\hline
    UNet~\cite{unet} &image  & 0.745 & 0.636 & 0.777 & 0.043 \\
    UNet++~\cite{unet++}  &image  & 0.749 & 0.633 & 0.780 & 0.039  \\
    TransUNet~\cite{chen2021transunet} &image  & 0.733 & 0.637 & 0.784 & 0.042 \\
    SETR~\cite{SETR} &image  &  0.709 & 0.588 & 0.748 & 0.045 \\
\hline
    STM~\cite{Oh_2019_ICCV_stm} & video & 0.741 & 0.634 & 0.782 & 0.041 \\
    AFB-URR~\cite{AFB-URR} & video & 0.750 & 0.635 & 0.781 & 0.038 \\
    PNS+~\cite{pns+} & video & 0.754 & 0.648 & 0.783 & 0.036 \\
    % GSFM~\cite{gsfm} &  &  &  &  & \\
    DPSTT~\cite{RethinkingLiJialu} &video & 0.755 & 0.649  & 0.785 & 0.036 \\
    DCFNet~\cite{dcfnet} & video & 0.762 & 0.659 & 0.799 & 0.037 \\
\hline
  Our FLA-Net & video & \textbf{0.789} & \textbf{0.687} & \textbf{0.815} & \textbf{0.033} \\
\hline 
\end{tabular}
\end{table}

\subsection{Comparisons with State-of-the-arts}
We conduct a comparative analysis between our network and nine state-of-the-art methods, comprising four image-based methods and five video-based methods.
Four image-based methods are UNet~\cite{unet}, UNet++~\cite{unet++}, TransUNet~\cite{chen2021transunet}, and SETR~\cite{SETR}, while five video-based methods are STM~\cite{Oh_2019_ICCV_stm}, AFB-URR~\cite{AFB-URR}, PNS+~\cite{pns+}, DPSTT~\cite{RethinkingLiJialu}, and DCFNet~\cite{dcfnet}.
%%%
To ensure a fair and equitable comparison, we acquire the segmentation results of all nine compared methods by utilizing either their publicly available implementations or by implementing them ourselves. Additionally, we retrain these networks on our dataset and fine-tune their network parameters to attain their optimal segmentation performance, enabling accurate and meaningful comparisons.

\vspace{3pt}
\noindent
\textbf{Quantitative Comparisons.}
The quantitative results of our network and the nine compared breast lesion segmentation methods are summarized in Table~\ref{tab:sota}. Analysis of the results reveals that, in terms of quantitative metrics, video-based methods generally outperform image-based methods. 
Among nine compared methods, DCFNet~\cite{dcfnet} achieves the largest Dice, Jaccard, and F1-score results, while PNS+~\cite{pns+} and DPSTT~\cite{RethinkingLiJialu} have the smallest MAE score.
%%%%%%
More importantly, our FLA-Net further outperforms DCFNet~\cite{dcfnet} in terms of Dice, Jaccard, and F1-score metrics, and has a superior MAE performance over PNS+~\cite{pns+} and DPSTT~\cite{RethinkingLiJialu}.
Specifically, our FLA-Net improves the Dice score from 0.762 to 0.789, the Jaccard score from 0.659 to 0.687, the F1-score result from 0.799 to 0.815, and the MAE score from 0.036 to 0.033.

\begin{figure*}[!t]
\centering
\includegraphics[width=0.90\textwidth]{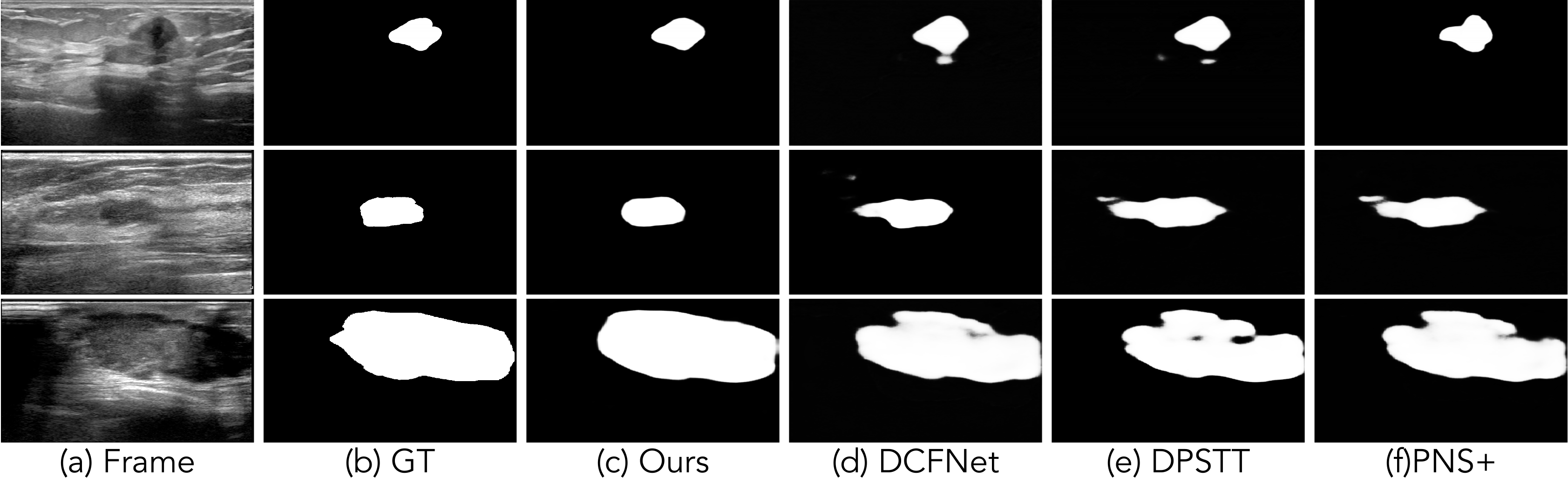}
\vspace{-2mm}
\caption{
Visual comparisons of breast lesion segmentation results produced by our network and state-of-the-art methods. ``GT'' denotes the ground truth. For more visualization results, please refer to the \textit{supplementary material}.
}
\label{fig:sota_cmp}
\end{figure*}

\begin{table}[!t]
\caption{
 Quantitative comparison results of ablation study experiments.
}
\vspace{-3mm}
\label{tab:ablation}
\centering
\resizebox{\columnwidth}{!}{
\begin{tabular}{l|ccc|c|c|c|c} 
  \hline
   & FLA & Loc-Branch &Contrastive loss & Dice $\uparrow$ & Jaccard $\uparrow$ & F1-score $\uparrow$& MAE $\downarrow$ \\
\hline
    Basic      &$\times$&$\times$&$\times$& 0.747 & 0.641 & 0.777 & 0.037   \\ 
    Basic+FLA  &$\checkmark$&$\times$&$\times$& 0.777 & 0.669 & 0.806 & 0.035  \\ 
    Basic+LB   &$\times$&$\checkmark$&$\times$& 0.751 & 0.646 & 0.781 &  0.037   \\
%\hline
   Basic+FLA+LB & $\checkmark$ & $\checkmark$ & $\times$ & 0.780 & 0.675 & 0.809 & 0.034   \\
\hline
    Our method & $\checkmark$ & $\checkmark$ & $\checkmark$ & \textbf{0.789} & \textbf{0.687} & \textbf{0.815} & \textbf{0.033}   \\
\hline
\end{tabular}
}
\end{table}

\vspace{3pt}
\noindent
\textbf{Qualitative Comparisons.}
Figure~\ref{fig:sota_cmp} visually presents a comparison of breast lesion segmentation results obtained from our network and three other methods across various input video frames. 
Apparently, our method accurately segments breast lesions of the input ultrasound video frames, although these target breast lesions have varied sizes and diverse shapes in the input video frames.

%%%%%%%%%%%%

\subsection{Ablation Study}
To evaluate the effectiveness of the major components in our network, we constructed three baseline networks. 
The first one (denoted as ``Basic'') removed the localization encoder branch and replaced our FLA modules with a simple feature concatenation and a 1$\times$1 convolutional layer. 
The second and third baseline networks (named ``Basic+FLA'' and ``Basic+LB'') incorporate the FLA module and the localization branch into the basic network, respectively.
Table~\ref{tab:ablation} reports the quantitative results of our method and three baseline networks.
The superior metric performance of "Basic+FLA" and "Basic+LB" compared to "Basic" clearly indicates that our FLA module and the localization encoder branch effectively enhance the breast lesion segmentation performance in ultrasound videos.
Then, the superior performance of ``Basic+FLA+LB'' over ``Basic+FLA'' and ``Basic+LB'' demonstrate that combining our FLA module and the localization encoder branch can incur a more accurate segmentation result.
Moreover, our method has larger Dice, Jaccard, F1-score results and a smaller MAE result than ``Basic+FLA+LB'', which shows that our location-based contrastive loss has its contribution to the success of our video breast lesion segmentation method.

%%%%%%%%%%%%
\begin{table}[!t]
\caption{
  Quantitative comparison results on different video polyp segmentation datasets.
 For more quantitative results please refer to the supplementary material.
}
\label{tab:vps-sota}
\vspace{-3mm}
\centering
\resizebox{\columnwidth}{!}{
\begin{tabular}{cc|cccccc|c} 
  \hline
   & Metrics & UNet~\cite{unet} & UNet++~\cite{unet++}& ResUNet ~\cite{resunet}& ACSNet~\cite{ACSNet} & PraNet~\cite{pranet} & PNSNet~\cite{PNSNet} & Ours \\
\hline
    \multirow{5}{*}{\rotatebox{90}{CVC-300-TV}} 
    & Dice $\uparrow$ & 0.639 &0.649 &0.535 & 0.738 & 0.739 & 0.840 & \textbf{0.874} \\
    & IoU $\uparrow$          &0.525 &0.539 &0.412 & 0.632 & 0.645 & 0.745 & \textbf{0.789} \\
    & $S_{\alpha}$ $\uparrow$  &0.793 &0.796 &0.703 & 0.837 & 0.833 & \textbf{0.909} & 0.907 \\
    & $E_{\phi}$  $\uparrow$  &0.826 &0.831 &0.718 & 0.871 & 0.852 & 0.921 & \textbf{0.969} \\
    & MAE      $\downarrow$     &0.027 &0.024 &0.052 & 0.016 & 0.016 & 0.013 & \textbf{0.010} \\
\hline
    \multirow{5}{*}{\rotatebox{90}{CVC-612-V}} 
    & Dice  $\uparrow$        &0.725 &0.684 &0.752 & 0.804 & 0.869 & 0.873 & \textbf{0.885} \\
    & IoU   $\uparrow$        &0.610 &0.570 &0.648 & 0.712 & 0.799 & 0.800 & \textbf{0.814} \\
    & $S_{\alpha}$ $\uparrow$ &0.826 &0.805 &0.829 & 0.847 & 0.915 & \textbf{0.923} & 0.920 \\
    & $E_{\phi}$ $\uparrow$   &0.855 &0.830 &0.877 & 0.887 & 0.936 & 0.944 & \textbf{0.963} \\
    & MAE   $\downarrow$      &0.023 &0.025 &0.023 & 0.054 & 0.013 & 0.012 & \textbf{0.012} \\
\hline 
\end{tabular}
}
\end{table}

\subsection{Generalizability of our network}
To further evaluate the effectiveness of our FLA-Net, we extend its application to the task of video polyp segmentation.
%%%%%%
Following the experimental protocol employed in a recent study on video polyp segmentation~\cite{PNSNet}, we retrain our network and present quantitative results on two benchmark datasets, namely CVC-300-TV~\cite{CVC-300} and CVC-612-V~\cite{CVC-612}.
Table~\ref{tab:vps-sota} showcases the Dice, IoU, $S_{\alpha}$, $E_{\phi}$, and MAE results achieved by our network in comparison to state-of-the-art methods on these two datasets.
Our method demonstrates clear superiority over state-of-the-art methods in terms of Dice, IoU, $E_{\phi}$, and MAE on both the CVC-300-TV and CVC-612-V datasets.
Specifically, our method enhances the Dice score from 0.840 to 0.874, the IoU score from 0.745 to 0.789, the $E_{\phi}$ score from 0.921 to 0.969, and reduces the MAE score from 0.013 to 0.010 for the CVC-300-TV dataset.
Similarly, for the CVC-612-V dataset, our method achieves improvements of 0.012, 0.014, 0.019, and 0 in Dice, IoU, $E_{\phi}$, and MAE scores, respectively.
Although our $S_{\alpha}$ results (0.907 on CVC-300-TV and 0.920 on CVC-612-V) take the 2nd rank, they are very close to the best $S_{\alpha}$ results, which are 0.909 on CVC-300-TV and 0.923 on CVC-612-V.
Hence, the superior metric results obtained by our network clearly demonstrate its ability to accurately segment polyp regions more effectively than state-of-the-art video polyp segmentation methods.

% #############################
\section{Conclusion}
In this study, we introduce a novel approach for segmenting breast lesions in ultrasound videos, leveraging a larger dataset consisting of 572 videos containing a total of 34,300 annotated frames.
We introduce a frequency and location feature aggregation network that incorporates frequency-based temporal feature learning, an auxiliary prediction of breast lesion location, and a location-based contrastive loss. 
Our proposed method surpasses existing state-of-the-art techniques in terms of performance on our annotated dataset as well as two publicly available video polyp segmentation datasets. These outcomes serve as compelling evidence for the effectiveness of our approach in achieving accurate breast lesion segmentation in ultrasound videos.

\small {\noindent\textbf{Acknowledgments:} This research is supported by Guangzhou Municipal Science and Technology Project (Grant No. 2023A03J0671),  the Regional Joint Fund of Guangdong (Guangdong–Hong Kong–Macao Research Team Project) under Grant 2021B1515130003, the National Research Foundation, Singapore under its AI Singapore Programme (AISG Award No: AISG2-TC-2021-003), A*STAR AME Programmatic Funding Scheme Under Project A20H4b0141, and A*STAR Central Research Fund. }

\bibliographystyle{splncs04}
\bibliography{paper495.bib}
\end{document}